\documentclass[3p,twocolumn]{elsarticle}

\usepackage{amssymb}
\usepackage{amsmath}
\usepackage{lineno}
\journal{Chaos Solitons and Fractals}

\begin{document}


\begin{frontmatter}

\title{Competition among alliances of different sizes}

\author[label1]{Breno F. de Oliveira}

\author[label2]{Attila Szolnoki}
\ead{szolnoki.attila@energia.mta.hu}

\address[label1]{Departamento de F\'\i sica, Universidade Estadual de Maring\'a, 87020-900 Maring\'a, PR, Brazil}

\address[label2]{Institute of Technical Physics and Materials Science, Centre for Energy Research, P.O. Box 49, H-1525 Budapest, Hungary}

\begin{abstract}
To understand the biodiversity of an ecosystem cannot be understood by solely analyzing the pair relations of competing species. Instead, we should consider multi-point interactions because the presence of a third party could change the original microscopic outcome significantly. In this way an alliance may emerge where species, who may have biased relations otherwise, can protect each other from an external invader. Such an alliance can be formed by two, three or even more species. By introducing a minimal model where six species compete for space we here study how the size of an alliance determines the vitality of a formation. We show that in the majority of parameter space the group of the smallest size prevails and other solutions can only be observed in a limited parameter range. These phases are separated by discontinuous phase transitions which can only be identified by intensive numerical efforts due to serious finite size effects and long relaxation processes.
\end{abstract}


\end{frontmatter}

\section{Introduction}
\label{intro}

It is a long standing intellectual challenge to identify the key mechanisms which maintain biodiversity in an ecosystem formed by competing species \cite{rozenzweig_95,ives_s07,gracia-lazaro_srep18,szolnoki_jrsif14,swain_pnas22}. Beside its theoretical importance, however, such knowledge also has practical significance because an incapable human intervene into an ecological community may result in an undesired effect. We may think of cyclically dominated species, for instance, which is a widely accepted possibility to explain the enigma why not a single species survives due to an evolutionary process \cite{roman_jtb16,tainaka_ei21,park_c18,avelino_pre19b,nagatani_srep18,palombi_epjb20,vukov_pre13,park_c19c,szolnoki_epl20,yoshida_srep22}. Here, if we want to weaken a species by reducing its invasion rate towards its prey then the opposite system reaction can be observed and the stationary portion of the target species grows \cite{tainaka_pla95,avelino_epl21,liao_nc20}. This, frequently called as the survival of the weakest effect, is just a delicate example about the difficulty we face when trying to understand a many-member system where the interactions are characterized by a subtle graph.

Interestingly, the above mentioned cyclical dominance does not only keep all three competing species alive, but their mutual interdependence may provide a specific way to protect them against the invasion of an external species. If, for instance, the latter attacks a member of the loop, but the other member is a predator of the invader then the inner invasion can result in a situation when the external aggressor becomes a prey, hence the stability of the original group is maintained. In other words, the vicinity of a third party can reverse the direction of the invasion laid down by the original microscopic rule \cite{drescher_cb14,szolnoki_njp15}.
This mechanism can be considered as a defensive alliance which was studied intensively in the last two decades \cite{szabo_pre01b,kim_bj_pre05,szabo_pre07,mitarai_pre12, park_amc18,szolnoki_csf20b}. It is important to stress that beside the already mentioned rock-scissors-paper game, other strategies of evolutionary games may also show similar non-transitive interaction and therefore they can form conceptually identical coalitions \cite{szolnoki_epl15,guo_h_jrsif20,szolnoki_pre17,mao_yj_epl18,canova_jsp18,li_xp_pla20,szolnoki_pre10b}. Furthermore, the above described mechanism does not restricted to a three-member formation, but larger loops of a Lotka-Volterra-type system containing more participants may behave similarly \cite{avelino_pre20,park_c19b}. But we can also think on a smaller group where two, otherwise neutral, species may mutually protect each other from an external predator.

It is reasonable to extend this concept because not only a single species can fight against an alliance, but alliances can also compete with each other for space \cite{brown_pre19}. It is an important question whether we can identify specific features of an alliance which determine its vitality, hence the outcome of the mentioned fight can be predicted in general. To give an example, when two three-member alliances compete then the one where the inner cyclic dominance is more intensive may dominate the other one \cite{perc_pre07b}. On the other hand, the diversity of inner invasion rates could be a detrimental feature, because those alliance which is formed by equally strong members performs better than the one where the simultaneous presence of a weak and more aggressive partners form the group \cite{blahota_epl20}.

\begin{figure*}[t]
\centering
\includegraphics[width=5.5cm]{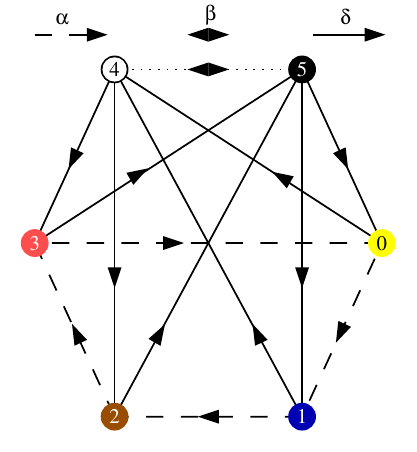}\hspace{2cm}\includegraphics[width=5.5cm]{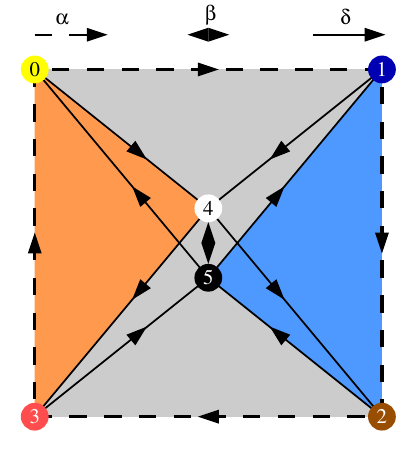}\\
\caption{Left panel shows the food-web of our six-species model where two different invasion rates are introduced, marked by $\alpha$ and $\delta$. Furthermore, the relation of species 4 and 5 is neutral, but they can exchange their next-nearest positions with probability $\beta$. The right panel shows an alternative representation of the same model, which helps to identify competing alliances more easily. The smallest coalition is formed by species 4 and 5 who may mutually protect their partner from an external invader. Three species, who dominate each other cyclically, can also from an alternative defensive partnership. For clarity we have marked these formations by colorized triangles. In particular, species 0, 4, and 3 and alternatively, species 1,2 and 5 have the mentioned rock-scissors-paper-like relation. Finally, species 0,1,2, and 3 at the corner of the grey box also form a four-species coalition, which may fight against the remaining two species effectively. This alternative representation of the food-web also highlights that the introduced invasion or site-exchange rates are strongly related to a specific type of alliances.}\label{foodweb}
\end{figure*}

In this work we consider another aspect and check whether the size of an alliance is a decisive factor in its vitality or not. For this goal we introduce a minimal six-species model where alliances with different sizes are present and compete for space. Perhaps it is worth noting that six-species models have already been studied previous, like in Refs.~\cite{szabo_jpa05,szabo_pre08b,esmaeili_pre18,baker_jtb20}, but our ecological system is designed to give a proper answer to the raised question. 

In our model the alliance specific invasion rates are the key parameters because, as we already noted, they could determine the strength of a specific formation. Our principal goal is to explore the multi-dimension parameter space and identify the dominant solution for each combination of control parameters. As we will show, the smallest alliance is the most effective defensive group in general and it prevails the majority of parameter space. Therefore larger formations composed by more participants can only win in specific ranges of control parameters. In the following we present some typical results which help us to identify some general observations. But before presenting them we proceed with a detailed description of the studied spatial multi-species model.

\section{A minimal six-species ecosystem}
\label{def}

In our spatial system six species, marked by $0, 1, \dots, 5$ indexes, are distributed randomly on a square lattice. No vacancies are allowed, all lattice sites are occupied by one of the species. To define the microscopic dynamics, we assume two types of interactions between nearest neighbors, which are invasion or site exchange. If, for example, species 4 and species 5 are nearest neighbors then they can exchange their positions with probability $\beta$. In case of the alternative interaction species 0 beats the neighboring species 1 with probability $\alpha$ and the empty site is occupied by an offspring of species 0. Similar relation is defined between species 1 and 2, between species 2 and 3, and finally between species 3 and 0 in cyclic manner. 

Beside the mentioned non-transitive interactions, we also declare predator-prey relations between species 4 (species 5) and the remaining species. For example, species 4 invades a neighboring species 2 or 3, but becomes the prey of a species 0 or 1. The mentioned invasions are executed with probability $\delta$. Similar relations can be defined for species 5 who invades a neighboring species 0 or 1, but becomes the prey of species 2 or 3.  
 
The microscopic rules are summarized in a food-web, shown in the left panel of Fig.~\ref{foodweb}, where the solid and dashed-line arrows depict a possible invasion while two sided arrows with dotted line mark the possibility of site exchange between specific species. The postulated invasion rates may seem to be artificial at first sight, but the alternative presentation of the same food-web, shown in the right panel of Fig.~\ref{foodweb}, reveals that they are strongly related to a certain type of alliances. In particular, the value of $\beta$ determines the site exchange between the neutral pair of 4 and 5 which establishes the smallest size of defensive alliance against an external species. For example, if species 0 attacks species 4 but the endangered species exchanges its position with a neighboring species 5 then the latter can repel the external invasion.

A three-party alliance is established among species 0, 4, and 3 who beats each other cyclically. If, for instance, species 1 attacks species 4 then a neighboring species 0 can neutralize this invasion. Similar rock-scissors-paper like alliance can be detected among species 1, 2, and 3. From the alternative presentation of the food-web it is also clear that the invasion rate $\delta$ characterizes partly the intensity of inner invasions in the three-size alliances. 

Last, a four-member alliance composed by the non-transitive interactions among species 0,1,2, and 3 can also be a promising fighter in our ecosystem. Here the intensity of the inner invasion is characterized by parameter $\alpha$. If, for instance, an external species 4 attacks species 2, who is a member of the larger group, then a neighboring species 1 can block the invasion of the intruder. Similarly, species 5 may also threaten the longest loop by attacking species 0 or 1, but neighboring species 2 or 3 becomes a guard in the mentioned situation.

According to the model design, we have alliances with different sizes which all can be a destination of an evolutionary process. Their relations, however, are far from obvious because we can always suggest a method how to defend the attack of an external species belonging to an adverse alliance. As we already mentioned, the power of a specific formation may depend sensitively on the inner dynamical processes, therefore the key parameters are $\alpha, \beta$, and $\delta$. While the first two parameters can be dedicated directly to the four-member and the two-member alliance respectively, the value of $\delta$ characterizes the interaction intensity between these formations. Therefore, one may expect that if $\delta$ is too small then there is no proper interaction between the mentioned alliances, hence there is no proper competition between them. As we will show later, this is a naive expectation.

Beyond the mentioned case, the parameter $\delta$ has another role in the microscopic dynamics. More precisely, the parameters $\alpha$ and $\delta$ together determines the inner flow in the three-member groups, which could affect the vitality of these formations. Summing up, there is no a clear preliminary expectation how our model should behave in dependence of the dynamical parameters.
Therefore, to answer our original question, we need to explore the whole three-dimensional parameter space and determine for every combination of $\alpha, \beta$, and $\delta$ values if there is a single victor or coexistence of alliances represents the stationary state.

Technically, we carried out Monte Carlo (MC) simulations on a square grid where the linear size was varied between $L=250$ and $5000$ lattice sites. We note that using the necessarily large system size, where the linear size of the square exceeds significantly the characteristic length size of the emerging patterns, is essential. Otherwise, a small-size simulation can easily result in misleading outcome which is not valid in the large-size limit. During an elementary step a player and its nearest neighbor are chosen randomly. If they represent different species then the possible elementary interaction is defined by the food-web shown in Fig.~\ref{foodweb}. This is executed by the given $\alpha, \beta$, or $\delta$ probabilities. If we repeat the described step $L \times L$ times then a full MC step is executed where on average all lattice sites have a chance to change. The necessary relaxation time to reach the final evolutionary destination were between $10^3$ and $10^6$ MC steps. To reach the expected accuracy we have averaged 100 independent runs starting from different initial states.

In the following we present our key observations about the battle of alliances composed by groups of different sizes.

\section{When the smallest is the strongest}
\label{results}

\begin{figure}[h!]
\centering
\includegraphics[width=7.0cm]{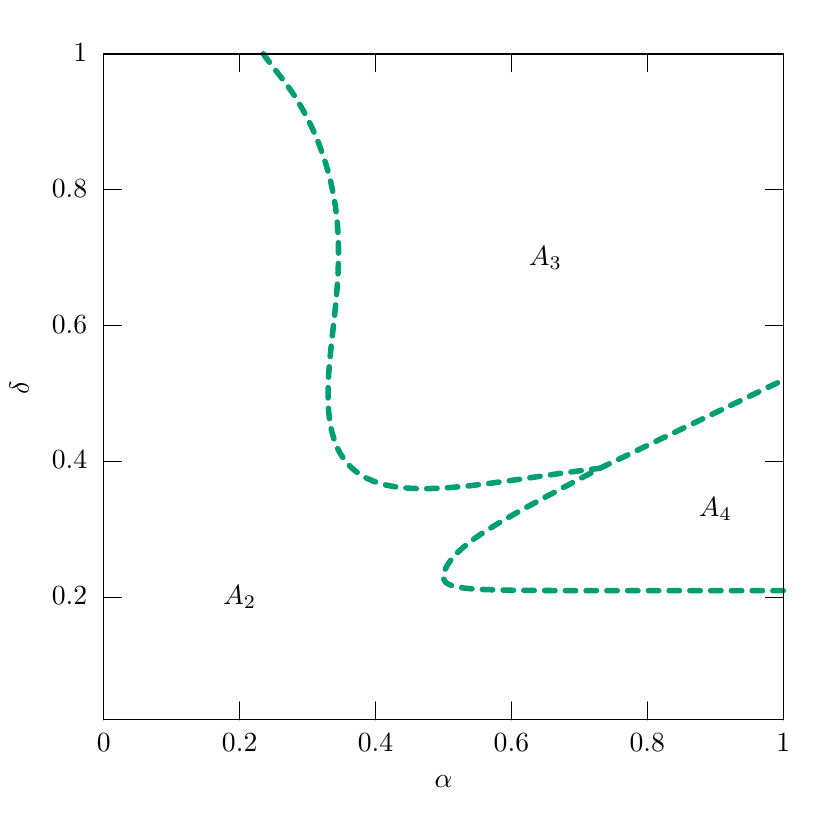}\\
\caption{Phase diagram on $\alpha - \delta$ plane obtained at fixed $\beta=0.01$ value. Here $A_i$ denotes the parameter region where the alliance containing $i$ members prevails. Note that only the very small $\beta$ value makes possible to observe alternative victors than the group of species $4+5$ who occupy $A_2$ domain. Dashed lines separating different phases mark discontinuous phase transitions.}\label{phd}
\end{figure}

If we consider the whole range of available parameter space, which is a $(0,1) \times (0,1) \times (0,1)$ cubic of probability values, then we can notice very fast that the alliance composed by the neutral pair of $4+5$ species is the most effective formation that dominates the majority of $(\alpha,\beta,\gamma)$ values. More precisely, if $\beta$, which controls the intensity of their site exchanges, is high enough then this alliance beats all the remaining groups independently of the values of $\alpha$ and $\delta$. The latter formations have only chance to win only when $\beta$ is very small. This case is illustrated in Fig.~\ref{phd} where we present the phase diagram obtained at a fixed $\beta=0.01$ value.

In this diagram we marked by $A_i$ the parameter region where the defensive group composed by $i$ species can win the battle of alliances. The general feature of this diagram confirms our expectations based on previous experiences about dynamically maintained defensive formations. Namely, the values of $\alpha$ and $\delta$ can be a decisive factor on who wins the competition. If $\alpha$ is small, for instance, then the four-party alliance simply has no chance to win: their inner invasion is too slow, hence they cannot block an external invasion effectively. More precisely, the appropriate member of the loop may arrive too late to the place where the external intruder attacks the territory of the four-member alliance. Therefore the latter group can only be viable if their inner flow is intensive, in other words $\alpha$ is high and $\delta$ is moderate simultaneously. The latter parameter ensures that species $4$ or $5$ cannot attack the remaining four species intensively. Indeed, the diagram of Fig.~\ref{phd} confirms our argument because the domain of $A_4$ can only be found in the large $\alpha$ - moderate $\delta$ region.

\begin{figure}
\centering
\includegraphics[width=7.5cm]{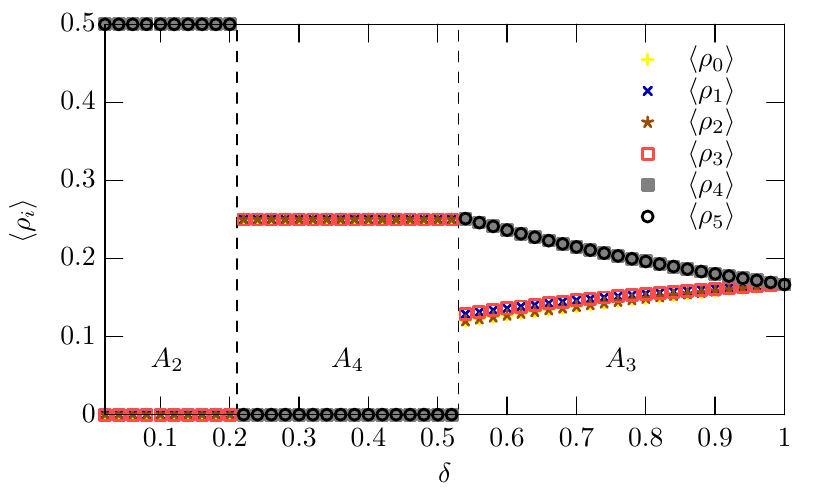}\\
\caption{The stationary concentrations of species in dependence of $\delta$ obtained at $\alpha=1, \beta=0.01$. We note that in the $A_3$ phase the $\rho_i$ values are the average of independent runs because here either $0+4+3$ or $1+2+5$ trio wins if we wait long enough. Accordingly, the concentration values for a specific destination are two times higher than those shown here.}\label{cross_a1}
\end{figure}

Before further discussing the phase diagram, we note that different phases are separated by sudden or more precisely discontinuous phase transitions. This behavior is illustrated by Fig.~\ref{cross_a1} where we made a cross section of the diagram by keeping $\alpha=1$ fixed. For small $\delta$ values $A_2$ prevails hence species $4$ and $5$ occupy the available space equally. One may argue that in case of small $\delta$, there is no proper interaction between the two- and the four-member alliances. Therefore they might coexist below a critical $\delta$ value. But this is not the case here because even a very weak interaction reveals the superiority of the species $4+5$ tandem. 

\begin{figure*}[t]
\centering
\includegraphics[width=15.0cm]{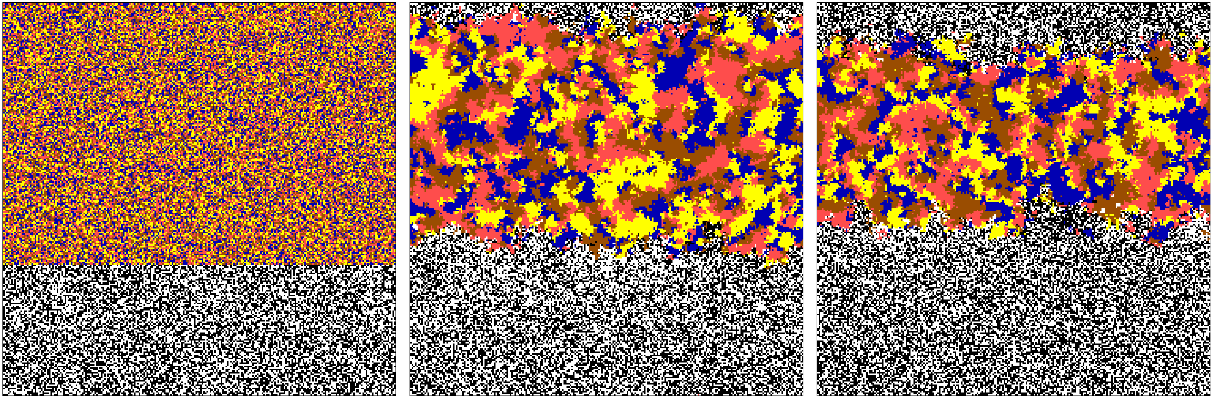}\\
\caption{The competition of $4+5$ and $0+1+2+3$ alliances at $\delta=0.0005$ when their interaction is extremely weak. The remaining parameters are $\alpha=1, \beta=0.01$. The linear size $L=250$ and the color coding of species is identical to the one we used in Fig.~\ref{foodweb}. In the initial state, shown in the left panel, $4+5$ species occupy one third portion of the available space randomly, while the other two thirds of space are occupied by the remaining four species. Both alliances emerge very fast and after they start to compete each other for space. Because of the small $\delta$ this process is very slow: second and third panels show the stage after 50,000 and 100,000 MC steps. But the tendency is clear and finally, not shown here, the alliance of $4+5$ species prevails.}\label{small_delta}
\end{figure*}

This process is illustrated clearly in Fig.~\ref{small_delta} where we monitor the competition of these solutions by applying an extremely small $\delta=0.0005$(!) invasion rate between the species belonging to different sets. At the beginning, shown in the first panel of Fig.~\ref{small_delta}, species $4+5$ are distributed randomly in the one third area of available space, while the other two thirds are filled by the remaining four species. When we launched the evolution both solutions emerge locally very fast. Namely, both the pair of $4+5$ and the quartet of $0+1+2+3$ form a solution which would be stable in the absence of the other formation. Eventually, however, the two-member alliance invades the other solution. Evidently, this process is very slow
because the interaction between the competing alliances is extremely weak: note that only 5 of 10000 meetings of competitor species would result in actual change at the frontiers! Therefore, not really surprisingly, after 100000 MC steps the $4+5$ tandem conquered just a small area from the $0+1+2+3$ quartet, as it is illustrated by the right panel in Fig.~\ref{small_delta}. But the tendency is clear. If we wait long enough, typically $10^6$ steps at $L=250$ system size, only species $4$ and $5$ survive. Perhaps it is worth mentioning that this final destination can be reached much faster if we launch the evolution from a completely random initial state where all six species are distributed uniformly, but here our principal goal was to illustrate the proper competition of two otherwise stable solutions. From this phenomenon we may conclude that an alliance having less members could be more effective than a larger coalition which is based on the collective acts of more members. 

\begin{figure*}
\centering
\includegraphics[width=15.0cm]{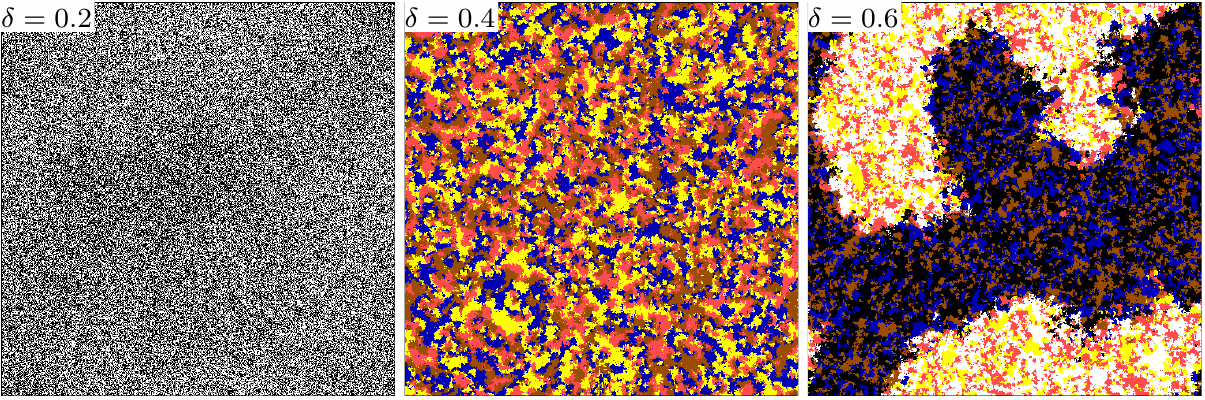}\\
\caption{Typical patterns obtained in different phases as we increase the value of $\delta$ when $\beta=0.01$ and $\alpha=1$. The color code of species are similar to we used in Fig.\ref{foodweb}. For small delta, only the two-species alliance survives. For intermediate value of $\delta$ the four-species formation prevails. In the range of large $\delta$ values the three-species loops beat all other alliances. Importantly, the right panel shows just an intermediate state of the evolutionary process because if we late long enough then one of the triplets will occupy the whole lattice.}\label{pattern_a1}
\end{figure*}

The relation of these alliances changes if we increase $\delta$ and the four-member group wins. Accordingly, all involved species occupies one fourth of the available space. The characteristic snapshots of $A_2$ and $A_4$ phases are shown in the first two panels of Fig.\ref{pattern_a1}. Interestingly, however, Fig.~\ref{cross_a1} suggests that by increasing $\delta$ further the triplet-based alliances can beat the four-member coalition. Based on the above mentioned conclusion one may ask that why  $\emph not$ an $A_3$ solution always dominates $A_4$?

The explanation is based on a previous work which studied the vitality of three-member alliances when heterogeneous inner invasion rates were used \cite{blahota_epl20}. As we already mentioned, a defective alliance based on cyclic dominance of the members may become sensitive if the inner invasion flow is too heterogeneous. In this case one of the members forms a large homogeneous domain, which makes the whole group vulnerable. If $\delta$ is too low compared to $\alpha=1$ then the trio group is not efficient and the quartet can win. Figure~\ref{cross_a1} demonstrates nicely that $A_3$ beats $A_4$ only above a threshold $\delta$ value. In the latter case $\delta$ value becomes comparable to $\alpha=1$ invasion rate resulting in a more balanced inner flow and a more uniform domain size distribution. Accordingly, a fit $A_3$ trio can already beat an $A_4$ quartet in agreement with our previously declared conclusion. The phase diagram shown in Fig.\ref{phd} also underlines that $A_3$ domain occupies the right-up corner of the phase space where $\alpha$ and $\delta$ values are comparable. But the unequal $\alpha$ and $\delta$ values result in a slightly different stationary concentrations of species, as it is illustrated in Fig.~\ref{cross_a1} \cite{tainaka_pla95}.
 
We would also like to stress that the spatial distribution of species shown in the right panel of Fig.~\ref{pattern_a1} does not represent the final destination of the evolutionary process. This plot only illustrates an intermediate stage when three-member alliances beat the alternative defensive groups. Indeed, the trio of $0+4+3$ and $1+2+3$ species are equally strong because both loops contain two $\alpha$ and a $\delta$ invasion rates, but they do no coexist permanently. Here, in the absence of symmetry breaking, both three-member formations can be the final victor of the evolution. But the coarsening of their domains is logarithmically slow in the lack of interfacial tension. This dynamical process is identical to the one previous observed for voter model \cite{cox_ap86,dornic_prl01}.

\begin{figure}
\centering
\includegraphics[width=7.5cm]{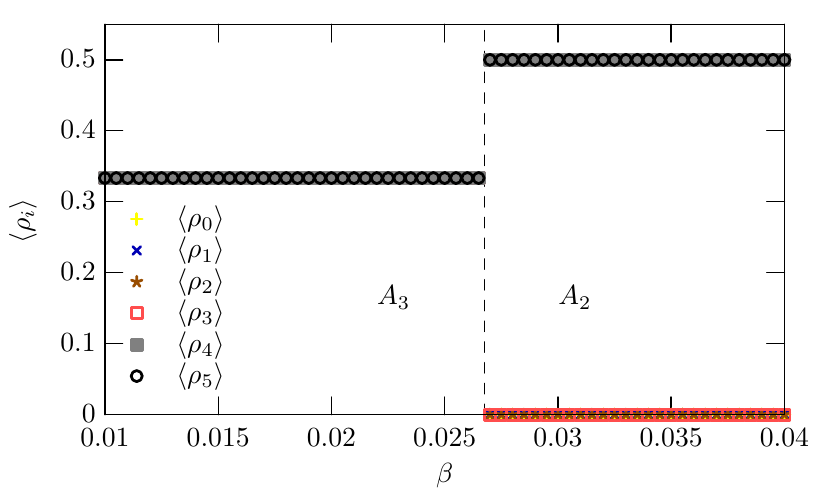}\\
\caption{The stationary concentrations of species in dependence of $\beta$ obtained at $\alpha=1, \delta=1$ values. If the site exchange between neutral partners becomes intensive the two-member alliance dominates the triplet formations. Note that in the $A_3$ phase either $0+4+3$ or $1+2+5$ triplet will win the competition.}\label{incr_b}
\end{figure}

Despite the diversity of phase diagram shown in Fig.~\ref{phd} we should not forget that the majority of parameter space is still dominated by the two-member $4+5$ solution. Their tandem is efficient almost everywhere if their site exchange capacity is not blocked. Put differently, if $\beta$ is high enough then they always win the evolutionary contest in our ecological system. To give an insight about the necessary intensity of site exchange we present a cross section starting from $A_3$ phase when we gradually increase the value of $\beta$. This is shown in Fig.~\ref{incr_b}. As we can see, if the intensity of site exchange is exceeds $\beta_c = 0.0270(2)$ critical value then the above mentioned duo gives no chance for alternative formations. Perhaps it is worth noting here that both $\alpha$ and $\delta$ parameter values are fixed, hence the microscopic details which determine the vitality of the triplets are intact. Therefore the sole change of $\beta$ influences the vitality of the neutral pair directly.

\begin{figure}[h!]
\centering
\includegraphics[width=7.5cm]{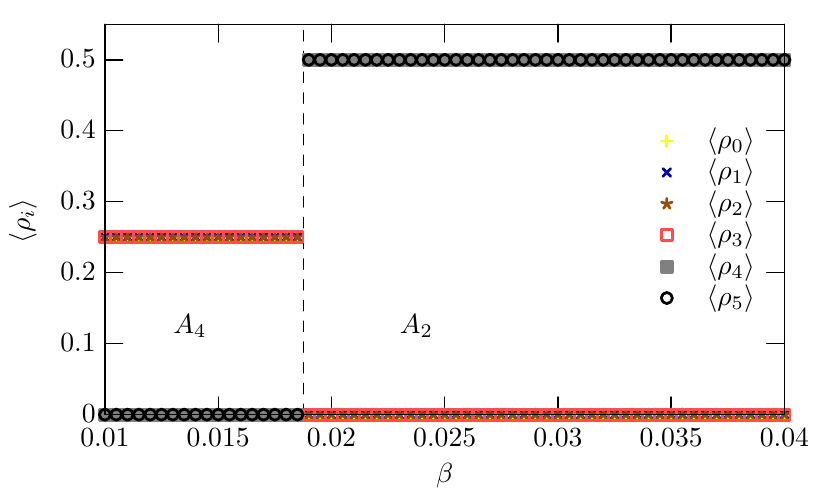}\\
\caption{The stationary concentrations of species in dependence of $\beta$ obtained at $\alpha=1, \delta=0.4$ values. The $0+1+2+3$ quartet can win the competition only if the site exchange between the neutral $4+5$ pair is extremely rare. Otherwise the latter formation will always win.}\label{incr_b_d}
\end{figure}

A similar behavior can be seen if we increase $\beta$ at a smaller $\delta=0.4$ value. As we already shown in Fig.~\ref{phd}, in this case the four-member loop of species 0, 1, 2, and 3 dominates the small $\beta$ limit. But their kingdom is fragile because already a small increase of $\beta$ can change this outcome. More precisely if the site exchange exceeds $\beta_c = 0.0190(2)$ threshold value then the neutral pair of $4+5$ will be the victor again.

\section{Conclusion}
\label{conclusion}

The success of network science in last two decades revealed the importance of interaction graphs both in ecological and conceptually similar alternative systems where the collective interaction of participants result in highly complex behavior \cite{szolnoki_epl11,amaral_pre20,tao_yw_epl21,zhu_pc_epjb21,amaral_pre21,kang_hw_pla21,quan_j_csf21,szolnoki_epl16,flores_jtb21,li_k_csf21,fu_mj_pa19}. A closed loop in a food-web could be specially interesting, because it offers a sort of higher level interaction when a group of participants, as an alliance, fight against an external player or a group \cite{cazaubiel_jtb17,garde_rsob20,szolnoki_srep16b}.

Our present work, however, highlights that not only the topology of the graph, but also the intensity of dynamical process along the available invasion path could be a decisive factor: we stress that the topology of food web is fixed, but the final evolutionary outcome could be very different depending on the invasion rates. We designed the graph to make possible for different alliances to emerge and our principal question to explore which of them is the final victor of the competition. As a general observation, we found that the alliance with the smallest size is the most effective in a huge area of parameter space and other competitors can win only in limited cases. More precisely, if the site exchange of neural partners exceed $\beta>0.03$ then their formation will always beat all other alliances independently of the values of the remaining parameters. This conclusion seems to be reasonable because in the mentioned optimal case only a single partner is needed to build a protective shield against others. According to this argument a defensive group formed by more members is always more vulnerable. But this is not necessarily the case because we presented situations when a four-member loop could be more efficient than a three-member group. This unexpected outcome reveals the importance of balanced inner invasions among the members of alliances. In this way we could provide further example and extend previously laid conjecture that the well-known rock-scissors-paper type defending alliance can only work properly if it is formed by fairly equal partners.   

We would like to note that our model is a minimal one because several, otherwise important, features are ignored here. For example, the migration or the mobility of species is another factor which could modify the stable solutions significantly \cite{reichenbach_n07,mobilia_g16,avelino_pre18,nagatani_jtb19,serrao_epjb21}. To extend our model with this feature promises further exciting observations because mobility may have unequal impact on a small or a large-size alliance.

\vspace{0.5cm}

B.F.O. thanks Funda\c c\~ao Arauc\'aria, and INCT-FCx (CNPq/FAPESP) for financial and computational support.

\bibliographystyle{elsarticle-num-names}

\end{document}